\documentclass[epj]{svjour}
\usepackage{graphicx}
\usepackage{multirow}
\usepackage{graphics}
\usepackage{hyperref}
\hypersetup{colorlinks = true, allcolors = blue}
\usepackage[utf8]{inputenc}
\usepackage{authblk}
\usepackage{hepparticles}
\usepackage{hepunits}
\usepackage{hepnames}

\newcommand{\epem}{\rm e^+e^-}

\begin{document}
\title{FCC-ee overview: new opportunities create new challenges}
\author{Alain Blondel\inst{1} \and Patrick Janot\inst{2} 
}                     
\offprints{}          
\institute{LPNHE, IN2P3/CNRS, Paris France, and University of Geneva, Switzerland \and CERN, EP Department, Geneva, Switzerland 
}
%

%
\abstract{
With its high luminosity, its clean experimental conditions, and a range of energies that cover the four heaviest particles known today, FCC-ee offers a wealth of physics possibilities, with high potential for discoveries. The FCC-ee is an essential and complementary step towards a 100\,TeV hadron collider, and as such offers a uniquely powerful combined physics program. This vision is the backbone of the 2020 European Strategy for Particle Physics. One of the main challenges is now to design experimental systems that can, demonstrably, fully exploit these extraordinary opportunities.
}

\date{{\sl (Submitted to EPJ+ special issue:
A future Higgs and Electroweak factory (FCC): Challenges towards discovery, Focus on FCC-ee)}}

%
%
\maketitle

\section{Introduction}
\label{section:intro}
The concept of a high-luminosity $\epem$ Higgs and Electroweak factory was born of a conjunction of opportunities in 2010-2012. Chronologically first, the objective of increasing the hadron collider energy by an order of magnitude up to ${\cal O}$(100\,TeV), combined with the limitation of high field magnets to less than 16-20 T, led to discussions of a new ring of 80-100\,km circumference. This range of circumference is luckily optimal given the geographical configuration of the relief around Geneva~\cite{todesco2012proceedings,ref:vhelhc1}. 

The first hints of a light Higgs boson in winter 2011~\cite{ATLAS:2012ae,Chatrchyan:2012tx}, followed by its discovery~\cite{Fabiola,Joe} in 2012 around 125\,GeV/$c^2$~\cite{HIGG-2012-27,CMS-HIG-12-028}, not much higher than the 95\% C.L. lower limit set by LEP ten years earlier, led to the investigation of a possible rebuild of an $\epem$ collider in the LEP/LHC tunnel. The circular collider technology had made great progress with the B factories (PEPII, KEKB, and the new Super-B~\cite{superbcollaboration2007superb} and SuperKEKB~\cite{10.1093/ptep/pts083} projects) and it had become possible to design a collider operating with a luminosity of more than $\rm 10^{34} cm^{-2} s^{-1} $ at the $\epem \to {\rm ZH}$ cross-section maximum in each of the several interaction regions~\cite{Blondel:2011fua}.      

Building a similar machine in a 100\,km tunnel~\cite{Blondel:2012ey} that would ultimately be hosting a 100\,TeV collider was seen as an interesting repeat of the LEP/LHC combination with its synergistic infrastructure and advantageous funding profile and, as would later be understood, fundamental physics complementarity. By a happy coincidence, a circumference of 80-100\,km is required for an $\epem$ circular collider to reach the top pair production threshold, enabling precise top quark measurements that are essential for the overall electroweak precision physics programme. Following the recommendations of the 2013 European Strategy for Particle Physics, the Future Circular Collider (FCC) conceptual design study was launched in February 2014, and comprised the study of both a hadron and a lepton collider.

The Conceptual Design Reports~\cite{Abada:2019lih,Benedikt:2651299,Benedikt:2018csr} were published by a large collaboration of 1500 authors. The overarching FCC document submitted to the European Strategy Update~\cite{Benedikt:2653673}, describes the FCC integrated programme (FCC-INT), a staged implementation scenario in which the lepton collider (FCC-ee) is the first step to be installed in a new 100 km underground circular infrastructure, providing a great part of the infrastructure towards the 100 TeV collider, as can be seen in Fig.\ref{fig:Schematic}. FCC-ee would be able to start taking data soon after the completion of the HL-LHC, for a duration of O(15 years). This would give time for the R\&D necessary to develop mass-produced and affordable 16-20T dipoles of accelerator-class quality and reliability; only then could the installation of the 100 TeV hadron collider (FCC-hh) be envisaged, during a ~10 years long shut down at the end of operation of FCC-ee. Cost estimates for i) each collider independently and ii) in the staged scenario, were given, demonstrating the considerable cost saving (7BCHF) realized in the FCC integrated scenario. Comparison with a number of alternative routes to a 100 TeV collider (low energy hadron collider or linear $\epem$ colliders) can be found in Ref.~\cite{Blondel:2019yqr} (see e.g. section 23) and led to the conclusion that FCC-INT provides the most efficient implementation of the FCC. The physics programmes of the two machines are both complementary and synergistic, and together offer a uniquely powerful long-term vision for high-energy physics.

This vision was adopted by the 2020 European Strategy for Particle Physics (ESPP)~\cite{CERN-ESU-015}: 
{\sl To prepare a Higgs factory, followed by a future hadron collider with sensitivity to energy scales an order of magnitude higher than those of the LHC}. 
The ESPP  placed the technical and financial feasibility of the Future Circular Colliders as the top priority (after the HL-LHC) for CERN and its international partners: {\sl Europe, together with its international partners, should investigate the technical and financial feasibility of a future hadron collider at CERN with a centre-of-mass energy of at least 100\,TeV and with an electron-positron Higgs and electroweak factory as a possible first stage. Such a feasibility study of the colliders and related infrastructure should be established as a global endeavour and be completed on the timescale of the next Strategy update.}

The CERN Council, in its June 2021 session, approved the FCC technical and financial feasibility study FCC-FS, with focus on the first step, i.e., the tunnel and the first stage machine FCC-ee. The study should deliver its report (FCC-FSR) end 2025. More details on the goals and milestones of the study, as well as on its organization  can be found in Refs.~\cite{FCC-FS-plans:202106} and \cite{FCC-FS-org:202106}. The R\&D on high field magnets for FCC-hh is treated independently, with high priority. Significant funding from CERN was approved for both activities. The FCC technical and financial feasibility study was launched at the FCC-week 28 June-2 July 2021~\cite{FCC-IS-week2021}.

\section{The FCC-ee collider}

\begin{figure}[ht]
\centering
\includegraphics[width=0.40\textwidth]{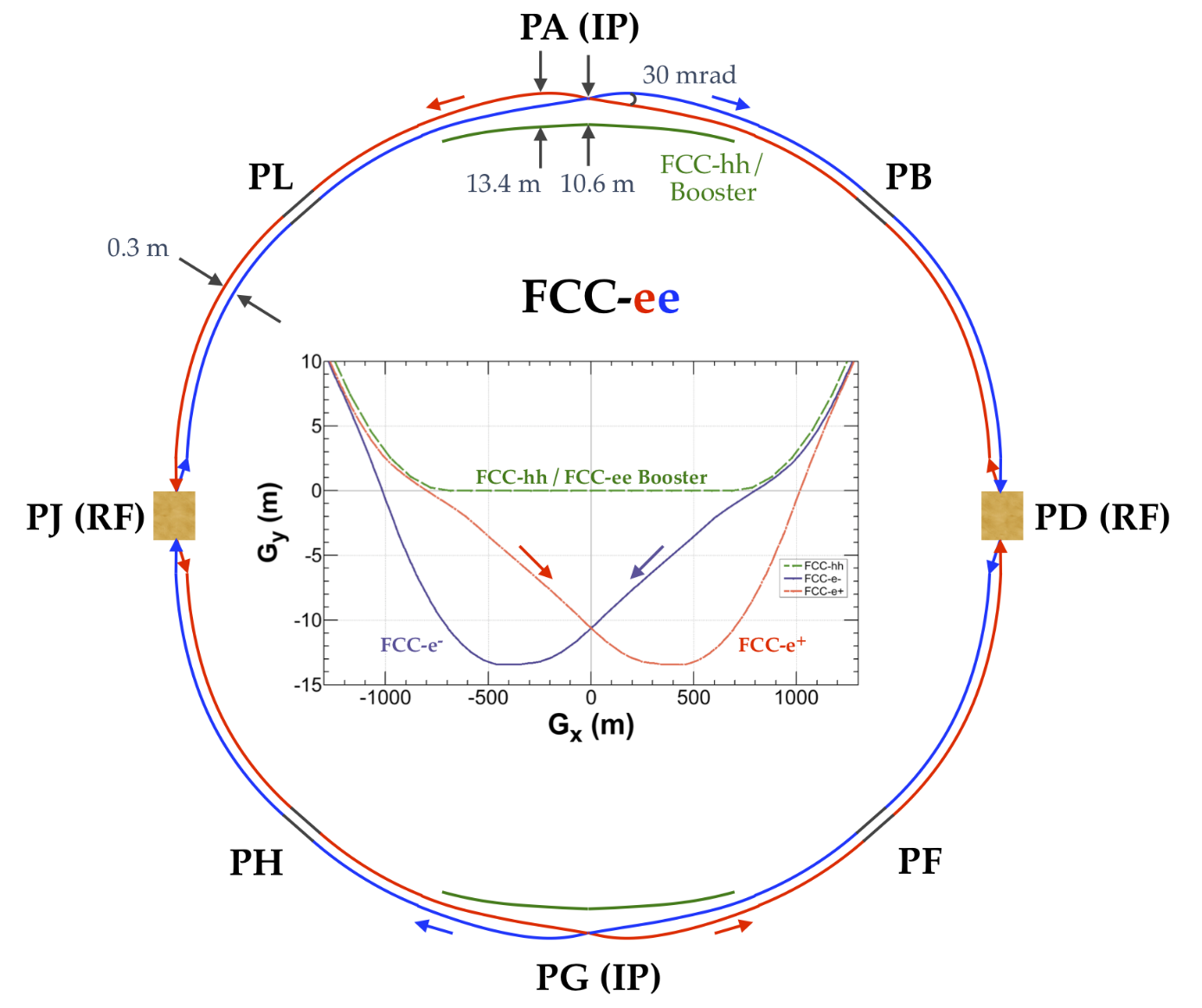}
\includegraphics[width=0.37\textwidth]{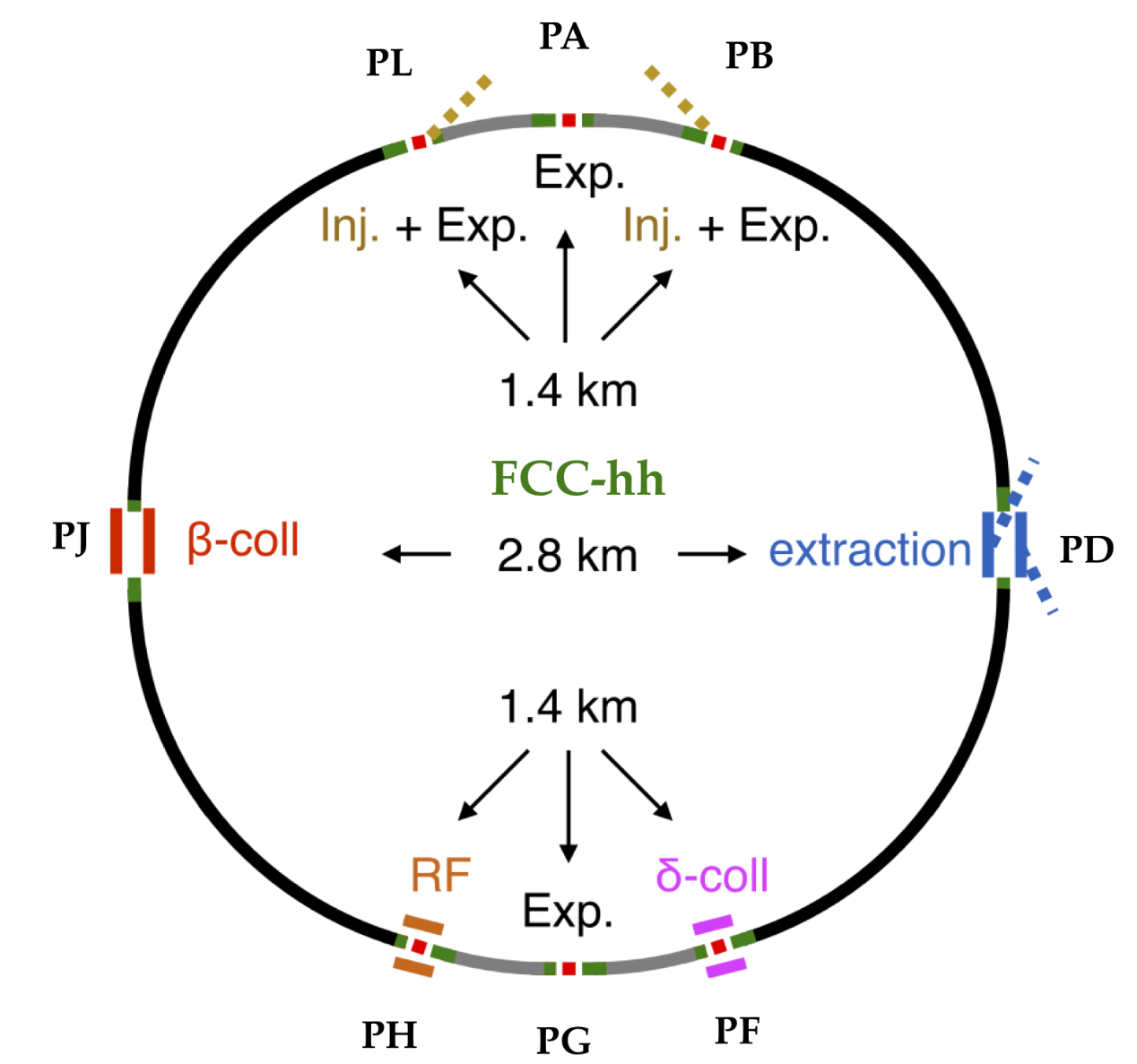}
\caption{\label{fig:Schematic} \small Schematics of the implementation of the FCC-ee collider (left) and the FCC-hh collider (right) in the common infrastructure~\cite{Benedikt:2651299}. The FCC-ee booster footprint coincides with that of the FCC-hh.~The asymmetric $\rm e^\pm$ beam lines around the FCC-ee interaction regions are designed to minimise synchrotron radiation in the detectors.}
\end{figure}
\begin{figure}[ht]
\centering
\includegraphics[width=0.60\textwidth]{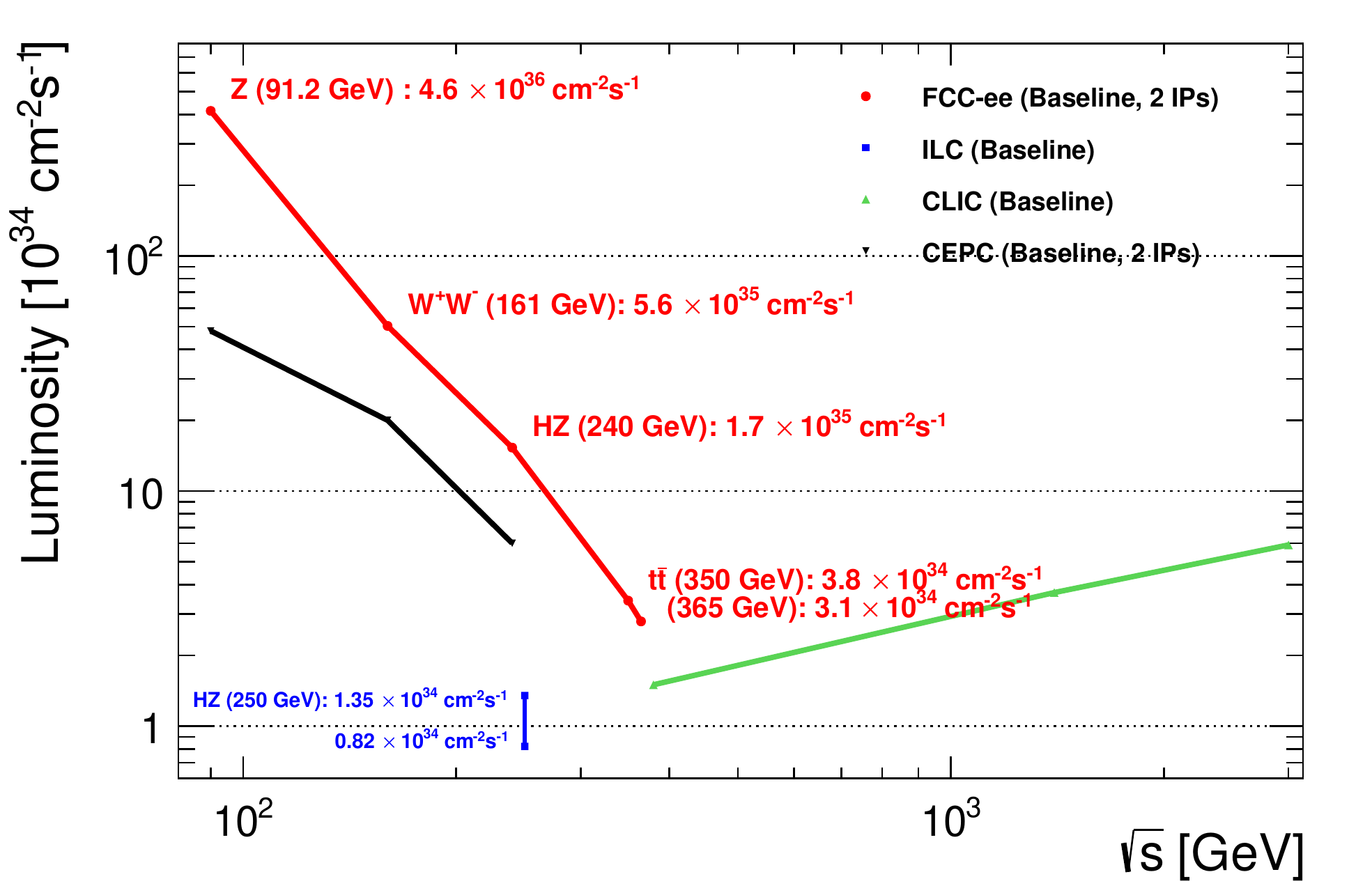}
\caption{\label{fig:OperationModel} \small The FCC-ee baseline design luminosity~\cite{Benedikt:2651299} summed over 2 IPs as a function of the centre-of-mass energy $\sqrt{s}$. Also indicated are the baseline luminosities of other ${\rm e^+e^-}$ collider proposals, as published in Refs.~\cite{Adolphsen:2013kya,Evans:2017rvt} (ILC), \cite{CLIC:2016zwp} (CLIC), and 
~\cite{CEPCStudyGroup:2018ghi} (CEPC).}
\end{figure}

As was already noted in a first study published in  2013~\cite{Gomez-Ceballos:2013zzn}, the FCC lepton collider expected performance is truly remarkable in the range of energies where lie the four high mass particles of the Standard Model. The design,  inspired by the B factories with top-up injection, strong focusing, and crab-waist optics, all fitting in the same tunnel as the subsequent 100\,TeV hadron collider as shown in Fig.~\ref{fig:Schematic}, maximises the luminosity. With two independent 100-km-circumference rings for $\rm e^+$ and $\rm e^-$, a total synchrotron power fixed to 100\,MW, and two interaction points (IPs),  a yearly integrated luminosity more than five orders of magnitude larger than that of LEP can be achieved at the Z pole, as shown in Fig.~\ref{fig:OperationModel}. The beam energy spread, which scales as $\sigma_{E_{\rm b}} \propto {E_{\rm b}}^2/\rho$ where $\rho$ is the ring bending radius~\cite{Sands:1969lzn}, is also reduced with respect to LEP, such that the natural build-up of transverse beam polarization provides conditions for a precise beam energy calibration up to above the W-pair energies. Moreover, the machine can operate over an energy range that extends up to and somewhat above the top-pair threshold. 

\vskip -.3cm

\vskip -.3cm

The FCC-ee will be implemented in stages as an {\bf electroweak, flavour, Higgs, and top factory} by spanning the energy range from the Z pole and the WW threshold through the maximum Higgs production rate, up to the ${\rm t\bar t}$ threshold and beyond. The high luminosity at the Z pole opens a unique domain for new particle searches~\cite{Blondel:2014bra}. The envisioned 15-year experimental programme is summarized in Table~\ref{tab:OperationModel}. The option to run on the Higgs resonance at $\sqrt{s}= m_{\rm H} = 125$\,GeV, as well as the possibility to operate four interaction points, are under investigation. 

\begin{table}[ht!]
\begin{center}
\caption{\small First five columns: The baseline FCC-ee operation model, showing the centre-of-mass energies, instantaneous luminosities for each IP, integrated luminosity per year summed over the 2 IPs corresponding to 185 days of physics per year and 75\% efficiency. As a conservative measure, the yearly integrated luminosity is further reduced by 10\% in this table and in all physics projections. The total luminosity distribution is set by the physics goals which in turn set the run time at each energy. The luminosity is assumed to be half the design value for commissioning new hardware during the first two years at the Z pole and in the first year at the ${\rm t\bar t}$ threshold. The sixth and last column is not part of the baseline FCC-ee operation model, but indicates possible numbers for an additional run at the H resonance, to investigate the electron Yukawa coupling. \vspace{0.4cm} \label{tab:OperationModel}}
\begin{tabular}{|l|c|c|c|c|c|c|c|}
\hline 
Working point & Z, {\footnotesize years 1-2} & Z, {\footnotesize later} & WW & HZ & \multicolumn{2}{|c|}{${\rm t\bar t}$  } & (s-channel H)  \\ \hline
$\sqrt{s}$ {\footnotesize (GeV)} & \multicolumn{2}{|c|}{88, 91, 94} & 157, 163 & 240 & {\small 340-350} & 365 & $\rm m_H $\\ \hline
{\small Lumi/IP {\footnotesize ($10^{34}\,{\rm cm}^{-2}{\rm s}^{-1}$)}} & 115 & 230 & 28 & 8.5 & 0.95 & 1.55 & (30) \\ \hline
{\small Lumi/year {\footnotesize (${\rm ab}^{-1}$, 2 IP)}} & 24 & 48 & 6 & 1.7 & 0.2 & 0.34 & (7) \\ \hline
Physics Goal {\footnotesize (${\rm ab}^{-1}$)} & \multicolumn{2}{|c|}{150} & 10 & 5 & 0.2 & 1.5& (20) \\ \hline
Run time {\footnotesize (year)} & 2 & 2 & 2 & 3 & 1 & 4 & (3) \\ \hline
 & \multicolumn{2}{|c|}{} & & $10^6$ \,HZ & \multicolumn{2}{|c|}{$10^6 {\rm t\bar t}$} & \\
Number of events &  \multicolumn{2}{|c|}{$5\times 10^{12}$ Z} & $10^8$ WW & + & \multicolumn{2}{|c|}{$+200$k HZ} & (6000) \\
 & \multicolumn{2}{|c|}{} & & 25k WW $\to$ H &  \multicolumn{2}{|c|}{$+50$k\,${\rm WW}\to {\rm H}$} &  \\ 
\hline
\end{tabular} 
\end{center}
\end{table}


\subsection{Centre-of-mass energy calibration} 
At the Z pole and the WW threshold, FCC-ee offers unparalleled control of the centre-of-mass energy and its distribution. The  procedures~\cite{Blondel:2019jmp} involve resonant depolarization, which provides a $\pm 2$~ppm instantaneous precision on the beam energies. It is complemented by a polarimeter, which also serves as beam spectrometer; and by extensive use of muon pairs produced in the experiments, to monitor with high precision the local difference between electron and positron energies, the beam crossing angle, and the centre-of-mass energy distribution. It was shown that, at the Z pole, the centre-of-mass energy scale should be known to 1\,ppm or better, with a point-to-point residual uncertainty of 40\,keV, and its spread to a couple per mil. The resulting precisions on the Z mass and width, the forward-backward asymmetry of muon pairs around the Z pole, as well as on the W mass, are listed in the first rows of Table~\ref{tab:EWPO}. 

These calibration procedures offer their own challenges~\cite{R0polblondel}, and are essential for the monitoring of the centre-of-mass energy for a possible run on the Higgs resonance at $\sqrt{s}= m_{\rm H}$. At the higher energies, the processes $\rm e^+ e^- \rightarrow Z\gamma$ and $\rm e^+ e^- \rightarrow WW$ and ZZ allow a determination of the centre-of-mass energy at the few MeV level, using the Z and W masses measured at the lower energies.

\subsection{The interaction region} 
\begin{figure}[htbp]
\centering
\includegraphics[width=0.50\textwidth]{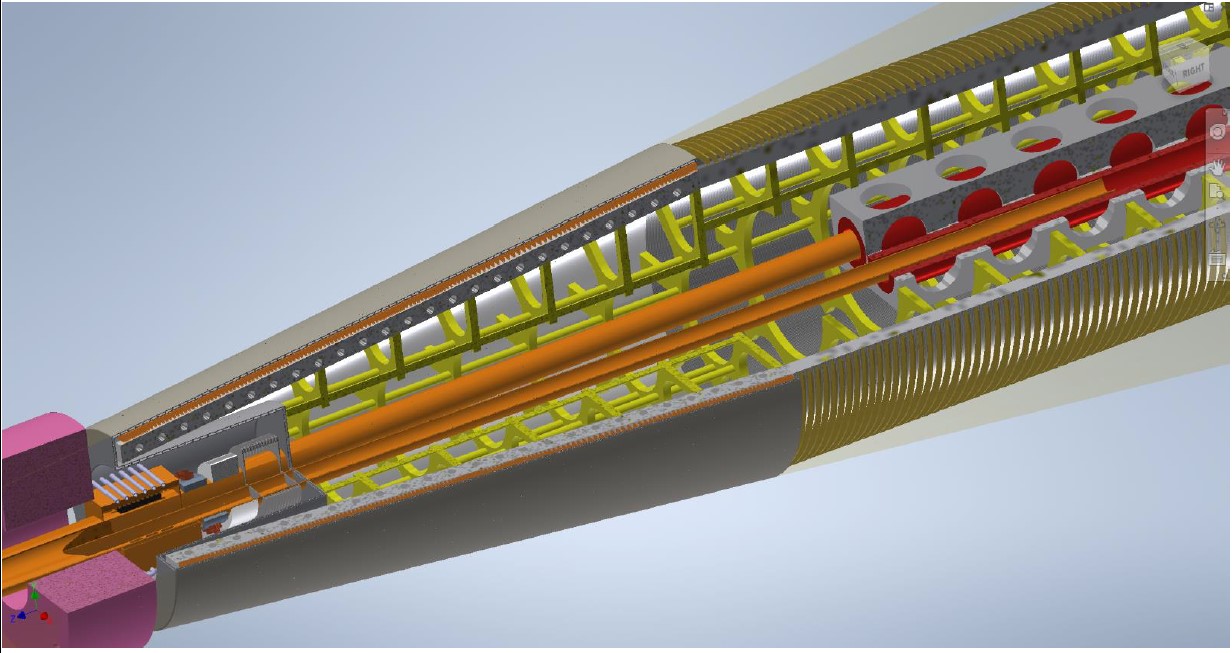}
\includegraphics[width=0.40\textwidth]{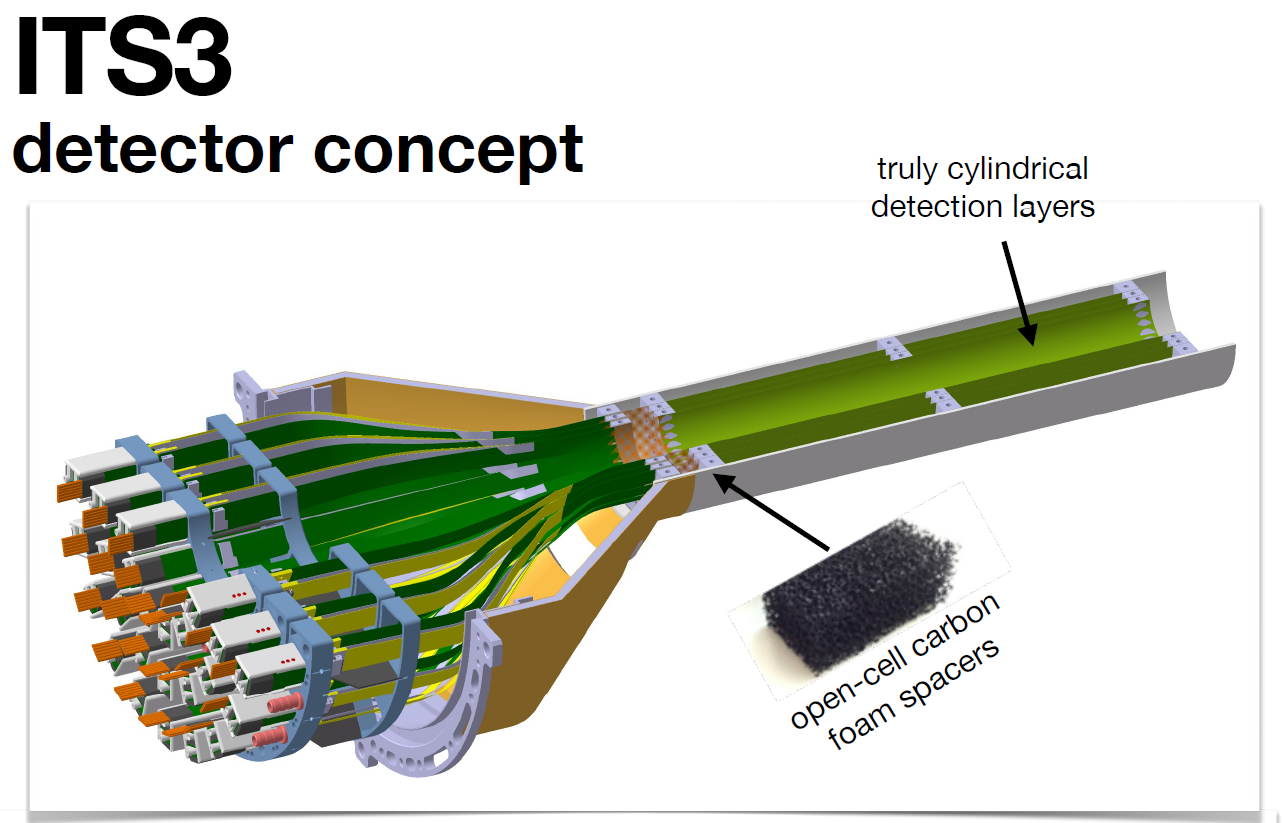}
\caption{
{\small Left: Layout of one side of the interaction  region; from left to right, the luminosity monitor (purple),  compensation solenoid and final focus quadrupoles, also showing the two beam pipes. Right: novel detector layouts will be studied for FCC-ee, similar to the ALICE ITS3 vertex detector project where cylindrical detection layers are situated inside the beam pipe~\cite{Mager:2747243}.}}
\label{fig:IR-magnets}
\end{figure}

The experimental environment at FCC-ee, where particles cross typically a few million times before interacting, is gentle and clean compared to hadron colliders, muon colliders and even linear $\epem$ colliders. The main aspects have been extensively studied by the Machine  Detector Interface group~\cite{Boscolo:2019awb}:
\begin{itemize}
    \item The design of the $\rm e^+$ and $\rm e^-$ rings is asymmetric around each interaction region, in order to eliminate strong bending magnets upstream of the collisions points, and  thus minimize the synchrotron radiation background in the detectors;
    \item The large number of bunches and the low level of beamstrahlung radiation result in a low rate of pile-up events, typically $2\times 10^{-3}$ at the Z pole, and less than $10^{-2}$ at the top energies; 
    \item The strong focusing optics requires a short free space (2.2 m) between the last beam elements. A compensating solenoid and final quadrupole assembly that fits in a forward dead cone of no more than 100~mrad, has been designed, as illustrated in Fig.~\ref{fig:IR-magnets}, left. The luminosity monitors situated in front of this assembly could be adapted to the geometry with two beam pipes crossing at an angle of 30\,mrad; 
    \item The small beam transverse dimensions allow for a beam pipe of 15\,mm radius. Recently, a radius of 10\,mm has been proposed and is being studied. The possibility of inserting the first layer of the vertex detector inside the vacuum chamber, as illustrated in the right panel of Fig.~\ref{fig:IR-magnets}, is also being discussed;
    \item Simulations indicate that the backgrounds due to beam-gas interactions and beamstrahlung lead to low occupancy;
    \item The detector solenoid magnetic field must be limited to 2\,T when operating at the Z pole, to avoid a blow up of the vertical beam emittance and a resulting loss of luminosity.  
\end{itemize}

These conditions are favourable on several accounts. The 100\,mrad low angle dead cone offers the possibility to build tracker and calorimeter down to a polar angle of $\cos{\theta}\simeq 0.99$, while the 10\,mm radius beam pipe should be a prime starting point for high efficiency b- and c-flavour tagging against light quarks and gluons. The 2\,T magnetic field limit is not a significant handicap since the momentum scale of the produced partons is typically distributed around 50\,GeV and does not exceed 182.5\,GeV. If needed, an increase of the magnetic field to 3\,T or more can be envisioned at higher energies, without loss of luminosity. 


\section{Opportunities: Higgs boson properties} 
\label{sec:Higgs}
The original motivation for an $\epem$ circular collider was to create a high-luminosity Higgs Factory~\cite{Azzi:2012yn,blondel2013report}, operating at the energy of the $\epem \to$ ZH cross-section maximum. This process is complemented by WW fusion to a Higgs boson, $\epem \to H \nu_e\bar\nu_e$, the cross section of which increases with the centre-of-mass energy (Table~\ref{tab:OperationModel}).
The total ZH cross section can be determined independently of the Higgs boson detailed properties by counting events with an identified Z boson, and for which the mass recoiling against the Z clusters around the Higgs boson mass~\cite{Azzurri:2021nmy}. This model-independent measurement of $g_{\rm HZZ}$, the coupling of the Higgs boson to the Z, is unique to $\epem$ colliders, and can be used as a ``standard candle'' by all other measurements, including those made at hadron colliders. The position of the recoil mass peak also provides an accurate measurement of the Higgs boson mass from the precise knowledge of the centre-of-mass energy. 
In combination with the measurement of the rate of HZ events with a $\rm H \to Z Z^\ast $ decay, proportional to $g_{\rm HZZ}^4/\Gamma_{\rm H}$, a model-independent determination of the Higgs total width $\Gamma_{\rm H}$ can be obtained. These numbers are impossible to extract from hadron collider data, for want of a well-defined initial state. The analysis of the other Higgs decays similarly provides a set of model-independent Higgs partial width and couplings. 

A summary of the expected FCC capabilities, combined (or not) with the HL-LHC projections, is displayed in Table~\ref{tab:kappaEFT}, for FCC-ee and for the full FCC-INT program. For the more curious reader, the Physics Briefing Book~\cite{Heinemann:2691414} offers a consistent analysis of other proposed colliders. 
The complementarity of the $\epem$ machine with  hadron colliders, and their synergies, are particularly well highlighted here. As soon as the Higgs coupling to the Z is known from the ZH total cross-section measurement, the LHC results for the relatively rare decays such as $\rm H\to \mu\mu, \gamma\gamma$ or Z$\gamma$, statistically more powerful than can be achieved with $\epem$ colliders alone, will become absolute measurements by normalising them to the  ${\rm H\rightarrow ZZ^\ast}$ decay. A similar comment applies to the top Yukawa coupling: it is determined with high statistical accuracy at a hadron collider (HL-LHC or FCC-hh), but must be normalized to the ttZ process to minimize otherwise dominant theory systematic uncertainties. The ttZ process uses in turn the FCC-ee measurement of the top electroweak couplings with $\epem \to t\bar{t}$ production. Thus not only does FCC-ee provide new model-independent measurements, it also renders more precise and model-independent those of the past and future hadron colliders. The same comment applies to the measurements at the FCC-ep collider, which provides additional statistical precision for several channels, in particular the the $\rm H \to WW^\ast$ and $\rm c \bar{c}$, but will benefit from the absolute normalisation from the FCC-ee measurements. 
{\setlength{\tabcolsep}{6pt} 
\renewcommand{\arraystretch}{1.} 
\begin{table}[!ht]
\centering

\caption{\small Precision on the Higgs boson couplings from Ref.~\protect\cite{deBlas:2019rxi}, in the $\kappa$ framework without (first numbers) and with (second numbers) HL-LHC projections, for the 
FCC-ee and the complete FCC integrated programme (including both the FCC-hh and the FCC-ep option). The HL-LHC result is obtained by fixing the total Higgs boson width and the ${\rm H \to c\bar c}$ branching fraction to their Standard Model values, and by assuming no BSM decays. All numbers are in \% and indicate 68\% C.L. sensitivities. Also indicated are the standalone precision on the total decay width, and the 95\% C.L. sensitivity on the "invisible" and "exotic" branching fractions. The precision on the Higgs self-coupling comprises the value extracted from the ZH and $\rm WW \to H$ cross sections for FCC-ee, with two (top line) or four (lower line) interaction points. The precision for FCC-hh has been updated using the most recent projections from double-Higgs production at FCC-hh~\cite{Mangano:2020sao}. } 
\label{tab:kappaEFT}

\vspace{2mm}
\begin{tabular}{|l|c|c|c|}
\hline Collider & {HL-LHC} &  
FCC-ee$_{240\to 365}$& FCC-INT \\ \hline
Lumi (${\rm ab}^{-1}$) &  3 &  
5 + 0.2 + 1.5&30 \\ \hline
Years & { 10} & 
3 + 1 + 4 & 25\\ \hline
$g_{\rm HZZ}$ (\%) &  1.5  & 
0.18 / 0.17 & 0.17/0.16 \\ 
$g_{\rm HWW}$ (\%) &  1.7  &
0.44 / 0.41 & 0.20/0.19 \\ 
$g_{\rm Hbb}$ (\%) &  5.1 & 
0.69 / 0.64 & 0.48/0.48\\ 
$g_{\rm Hcc}$ (\%) &  SM &
1.3 / 1.3 & 0.96/0.96\\ 
$g_{\rm Hgg}$ (\%) &  2.5  & 
1.0 / 0.89 & 0.52/0.5 \\ 
$g_{\rm H\tau\tau}$ (\%) &  1.9 & 
0.74 / 0.66 & 0.49/0.46 \\ 
$g_{\rm H\mu\mu}$ (\%) &  4.4  &
8.9 / 3.9 & 0.43/0.43 \\ 
$g_{\rm H\gamma\gamma}$ (\%) &  1.8  &
3.9 / 1.2 & 0.32/0.32 \\ 
$g_{\rm HZ\gamma}$ (\%) &  11.  & 
-- / 10.& 0.71/0.7 \\ 
$g_{\rm Htt}$ (\%) &  3.4 & 
10. / 3.1 & 1.0/0.95 \\ \hline
\multirow{2}{*}{$g_{\rm HHH}$ (\%)} &  \multirow{2}{*}{50.} &
44./33. & \multirow{2}{*}{3-4} \\ 
& & 
{27./24.}&  \\ \hline
$\Gamma_{\rm H}$ (\%) &  SM & 
1.1 & 0.91 \\ \hline
BR$_{\rm inv}$ (\%) &  1.9 & 
0.19 & 0.024\\ 
BR$_{\rm EXO}$ (\%) &  SM (0.0) &
1.1 & 1 \\ \hline
\end{tabular} 
\end{table}
}

As explained in Refs.~\cite{McCullough:2013rea,Maltoni:2018ttu,blondel2019future,Micco_2020}, the capability of FCC-ee to measure the ZH and $\rm WW \to H$ cross sections at two different centre-of-mass energies (240 and 365\,GeV) offers a 2-to-4 $\sigma$ sensitivity to the Higgs self-coupling, via a loop diagram. The FCC-ee Higgs and top coupling measurements make it possible to  measure the Higgs self coupling at the 100\,TeV collider, with a precision around  $\pm$ 10\% within a couple years running and eventually of ($\pm 3_{\rm stat} \pm 1.6_{\rm syst} )$\%~\cite{Mangano:2020sao}.
For the Higgs measurements as for many other aspects of the physics program, {\bf the combination of FCC-ee and FCC-hh is outstanding}~\cite{Heinemann:2691414}.  

Finally, the FCC-ee stands out among the  ``Higgs Factory" projects for its unique opportunity to access the Higgs boson coupling to electrons~\cite{Ghosh:2015gpa,Dery:2017axi,Altmannshofer:2015qra} through the resonant production process ${\rm e^+e^- \to H}$ at $\sqrt{s} = 125$\,GeV~\cite{dEnterria:2017dac}. This measurement relies on the combination of the high luminosity and monochromatization of the beams, and is under study. 
This is a unique opportunity of FCC-ee, and one of its toughest challenges.


\section{Opportunities: Precision measurements}

Several experimental facts require extensions of the Standard Model, in particular: the dominance of matter over antimatter in the Universe; the evidence for dark matter from astronomical and cosmological observations; and, closer to particle physics, the smallness of the neutrino masses, about $10^{-7}$ times smaller than that of the electron. 
The possible solutions to these questions seem to require the existence of new particles or phenomena that can occur over an immense range of mass scales and coupling strengths. 
The discovery of new particles, before their actual observation, has often been guided in the past 
by predictions based a long history of experiments and of theory maturation. In this context, a decisive improvement electroweak precision observable (EWPO) measurements, combined with other precision measurements of the properties of the Higgs boson, the top quark as well as b and c hadrons and the  tau lepton, could play a crucial role, by integrating sensitivity to a large range of new physics possibilities. The observation of significant deviation(s) from the Standard Model predictions would definitely be a discovery. A high significance requires not only a considerable improvement in experimental and theoretical precision, but also a large set of measured observables, in order to eliminate spurious deviations; and possibly reveal a pattern of deviations, enabling the guidance of theoretical interpretation. 
{\bf Improved precision equates discovery potential. It also challenges both theory and experiment}. Table~\ref{tab:EWPO} gives a good feel for the opportunities and the associated challenges presented by precision measurements. 



\setlength{\tabcolsep}{1.5pt}
\begin{table}[!ht]
\centering
\caption {Measurement of selected precision measurements at FCC-ee, compared with present precision. 
The systematic uncertainties are initial estimates, aim is to improve down to statistical errors.
This set of measurements, together with those of the Higgs properties, achieves indirect sensitivity to new physics up to a scale $\Lambda$
of 70\,TeV in a description with dim 6 operators, and possibly much higher in specific new physics (non-decoupling) models.  \vspace{1mm}\label{tab:EWPO}}
\begin{tabular}{|l|rcl|c|c|r|}
\hline
Observable  & present &  &          &  FCC-ee  &  FCC-ee  &  Comment and   \\
            & value  &$\pm$& error  &  {\bf Stat.} &   Syst.     & leading exp. error \\
\hline\hline
$ \mathrm{m_Z\,(keV) } $  &  91186700   & $\pm$ &  2200    & {\bf 4} & 100  & From Z line shape scan  \\
$  $  &  & &    &   &  & Beam energy calibration \\
\hline
$ \mathrm{  \Gamma_Z  ~(keV) } $  & 2495200   & $\pm$ &  2300    & {\bf 4}  &  25  & From Z line shape scan  \\
$  $  &  & &    &   &   &  Beam energy calibration  \\
\hline
$ \mathrm{ sin^2{\theta_{W}^{\rm eff}}} (\times 10^6) $  & 231480   & $\pm$ &  160   & {\bf 2}   &  2.4  & 
from $ \mathrm{ A_{\rm FB}^{{\mu} {\mu}}}$  at Z peak\\
$  $  &  & &    &   &   &  Beam energy calibration  \\
\hline
$ \mathrm{ 1/\alpha_{QED} (m_{\rm Z}^2) } (\times10^3) $  & 128952
  & $\pm$ &  14   & {\bf 3}   &  small  &
from $ \mathrm{ A_{\rm FB}^{{\mu} {\mu}}}$ off peak\\
$  $  &  & &    &   &  &  QED\&EW errors dominate  \\
\hline
$ \mathrm{  R_{\ell}^{Z}} ~(\times 10^3) $  & 20767 & $\pm$ &  25   & {\bf 0.06}   & 0.2-1   &  ratio of hadrons to leptons \\
$  $  &  & &    &   &   &  \bf acceptance for leptons  \\
\hline
$ \mathrm{ \alpha_{s} (m_Z^2) } ~(\times 10^4) $  &
 1196 & $\pm$ &  30  &  {\bf 0.1}  &  0.4-1.6  &
from $\mathrm{  R_{\ell}^{Z}}$ above\\
\hline

$ \mathrm{\sigma_{had}^0} ~(\times 10^3)$ (nb) & 41541 & $\pm$ &  37   & {\bf 0.1}  &  4  &  peak hadronic cross section  \\
$  $  &  & &    &   &   &  \small luminosity measurement  \\
$ \mathrm{  N_{\nu}}  (\times 10^3) $  & 2996  & $\pm$ &  7   & {\bf 0.005}   &  1  &  Z peak cross sections \\
$  $  &   &  &    &   &  &   Luminosity measurement \\
\hline
$ \mathrm{  R_b} ~(\times 10^6) $  & 216290 & $\pm$ &  660   & {\bf 0.3}    &  $<$ 60  &  ratio of $\rm{ b\bar{b}}$  to hadrons  \\
$  $  &  & &    &   &   &  stat. extrapol. from SLD
\\
\hline
$ \mathrm{  A_{\rm FB}^b,0} ~(\times 10^4) $  & 992 & $\pm$ &  16   & {\bf 0.02}   &  1-3  &  b-quark asymmetry at Z pole  \\
$  $  &  & &    &   &   &  from jet charge \\
\hline
$ \mathrm{A_{\rm FB}^{pol,\tau} ~(\times 10^4)} $  & 1498 & $\pm$ &  49  & {\bf 0.15}   &  $<$2  &  $\tau$ polarization asymmetry  \\
$  $  &  & &    &   &   &  $\tau$ decay physics \\
\hline
$\tau$ lifetime (fs)  &  290.3  & $\pm$ &  0.5   &  {\bf 0.001}   & 0.04   &  radial alignment  \\
\hline 
$\tau$ mass (MeV)  & 1776.86 & $\pm$ & 0.12  & {\bf 0.004} & 0.04   & momentum scale \\
\hline 
$\tau$ leptonic ($\mu \nu_\mu \nu_\tau$) B.R. (\%) & 17.38 & $\pm$ & 0.04  & {\bf 0.0001} & 0.003   & e/$\mu$/hadron separation \\ 
\hline 
$ \mathrm{ m_W  ~(MeV) } $  &  80350   & $\pm$ &  15    & {\bf 0.25}  & 0.3  & From WW threshold scan \\
$  $  &  & &    &   &  &  Beam energy calibration  \\
\hline
$ \mathrm{  \Gamma_W  ~(MeV) } $  & 2085   & $\pm$ &  42    & {\bf 1.2}  & 0.3  & From WW threshold scan \\
$  $  &  & &    &   &   &  Beam energy calibration  \\
\hline
$ \mathrm{ \alpha_{s} (m_W^2) }  (\times 10^4)$  &
 1170   & $\pm$ & 420 &  {\bf 3}  & small  &
  from $ \mathrm{R_{\ell}^{W} }$\\
\hline
$ \mathrm{  N_{\nu}}  (\times 10^3) $  & 2920 & $\pm$ &  50   & {\bf 0.8}   & small   &   ratio of invis. to leptonic \\
$  $  &  & &    &   &  & in radiative Z returns  \\
\hline
$ \mathrm{ m_{top}  ~(MeV/c^2) } $  &  172740   & $\pm$ &  500    & {\bf 17}  & small  & From $\mathrm {t\bar{t}}$ threshold scan \\
$  $  &  & &    &   &  &  QCD errors dominate  \\
\hline
$ \mathrm{ \Gamma_{top}  ~(MeV/c^2) } $  &  1410   & $\pm$ &  190    &{\bf  45}  & small  & From $\mathrm {t\bar{t}}$ threshold scan \\
$  $  &  & &    &   &  &  QCD errors dominate \\
\hline
$ \mathrm{ \lambda_{top}/\lambda_{top}^{SM}   } $  &   1.2
    & $\pm$ &  0.3    & {\bf 0.10}  & small  & From $\mathrm {t\bar{t}}$ threshold scan \\
$  $  &  & &    &   &  &  QCD errors dominate \\
\hline
$ \mathrm{ ttZ ~couplings   } $  &
   & $\pm$ &  30\%   & {\bf 0.5 -- 1.5} \%  & small  & From $\sqrt{s}=365$\,GeV run  \\
\hline
\end{tabular}
\end{table}

\begin{figure}[htbp]
\centering
\begin{minipage}[b]{0.495\textwidth}
\centering
\includegraphics[width=\textwidth]{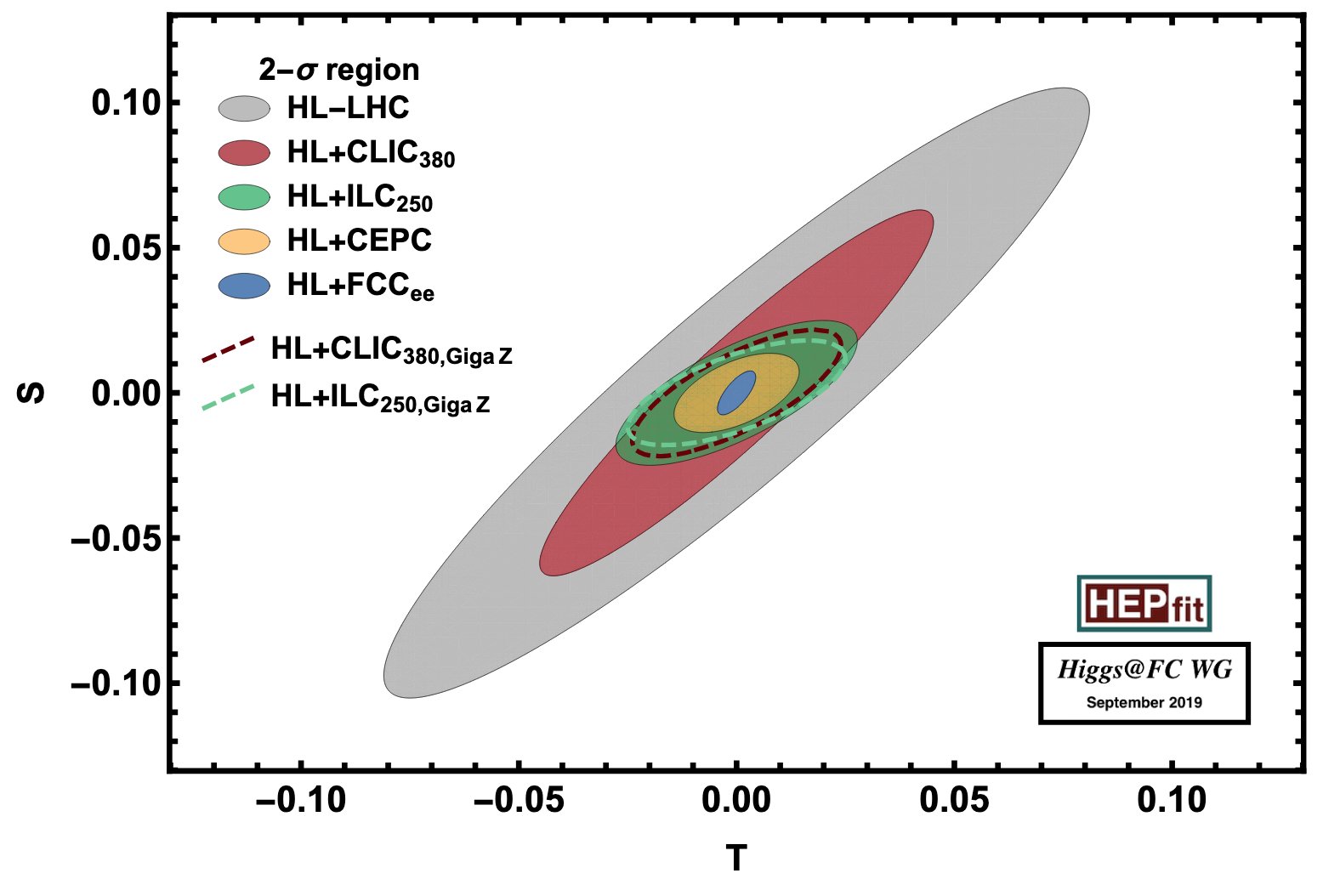}
\end{minipage}
\begin{minipage}[b]{0.482\textwidth}
\centering
\includegraphics[width=\textwidth]{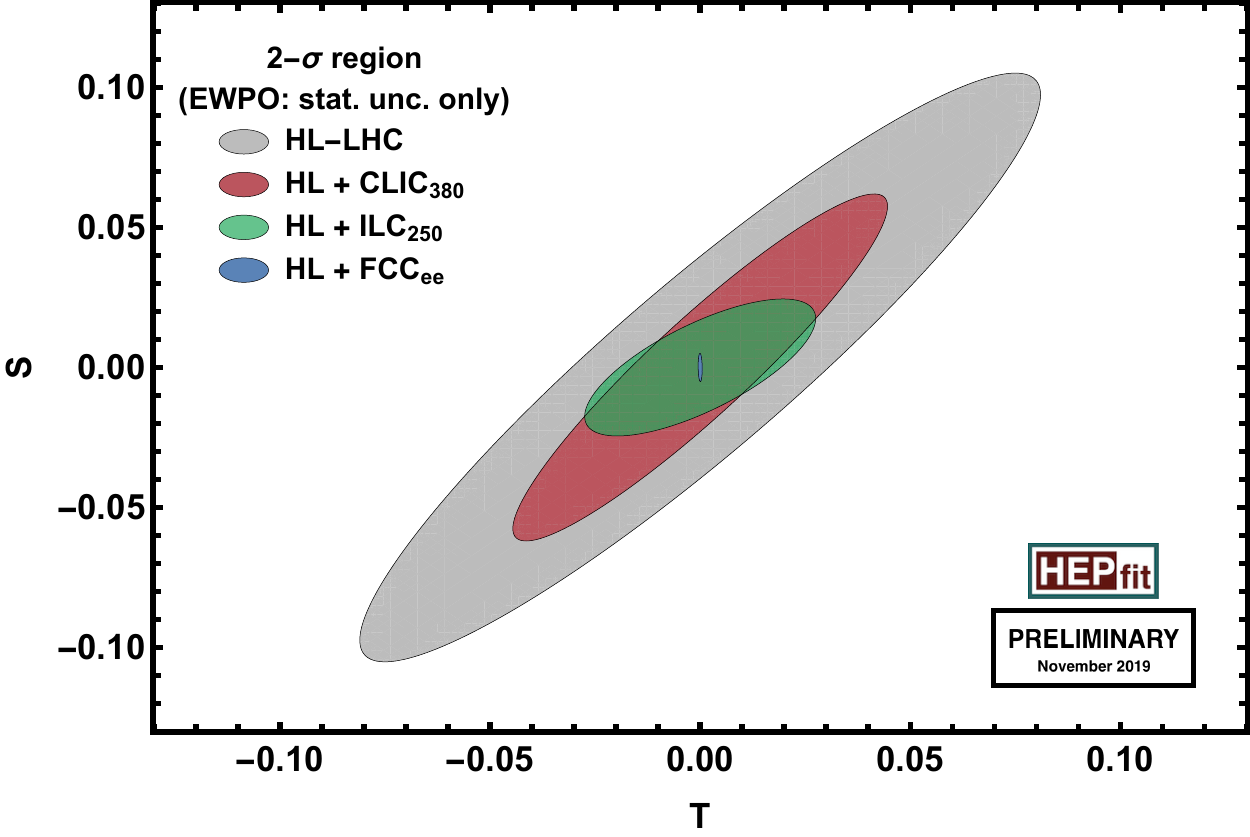}

\vspace{0.1cm}
\end{minipage}
\caption{\label{fig:EWPO} \small Expected uncertainty contour for the $S$ and $T$ parameters for various colliders in their first energy stage.  For ILC and CLIC, the projections are shown with and without dedicated running at the Z pole, with the current (somewhat arbitrary) estimate of future experimental and theoretical systematic uncertainty (left, from Ref.~\cite{Heinemann:2691414}); and with only statistical and parametric uncertainties (right, from Ref.~\cite{JdeBlas}).}
\end{figure}

The expected statistical precision for EWPOs measured at the Z pole is typically 500 times smaller than the current uncertainties. Some projections give much larger improvements even (e.g. $R_{\rm b}$). For the first time, a direct measurement of $\rm \alpha_{QED}(m_Z)$ will be possible~\cite{Janot:2015gjr} with ${\cal O} (10^{-5})$ precision. How far can one push on the experimental and theoretical systematic uncertainties? Can the statistical precision be matched? This perspective constitutes a superb opportunity and  considerable challenge for detector designers and for experts in theoretical calculations. 
As an illustration, Fig.~\ref{fig:EWPO} shows the projected uncertainty, for different ${\rm e^+e^-}$ colliders, on the $S$ and $T$ parameters (used to parameterise the isospin-conserving and isospin-breaking virtual effects in the Z and W propagators) from a global fit of EWPOs. The FCC-ee measurements at the Z pole, at the WW threshold, and at the ${\rm t\bar t}$ threshold, give the best perspective (left panel); and allow the parametric uncertainties to be reduced to a minimum (right panel). The right panel of Fig.~\ref{fig:EWPO} also shows the tremendous stand-alone true potential of FCC-ee, should future experimental and theoretical systematic uncertainties match the available statistics, a considerable challenge to the experiment and theory teams. 

\begin{figure}[htbp]
\centering
\includegraphics[width=0.80\textwidth]{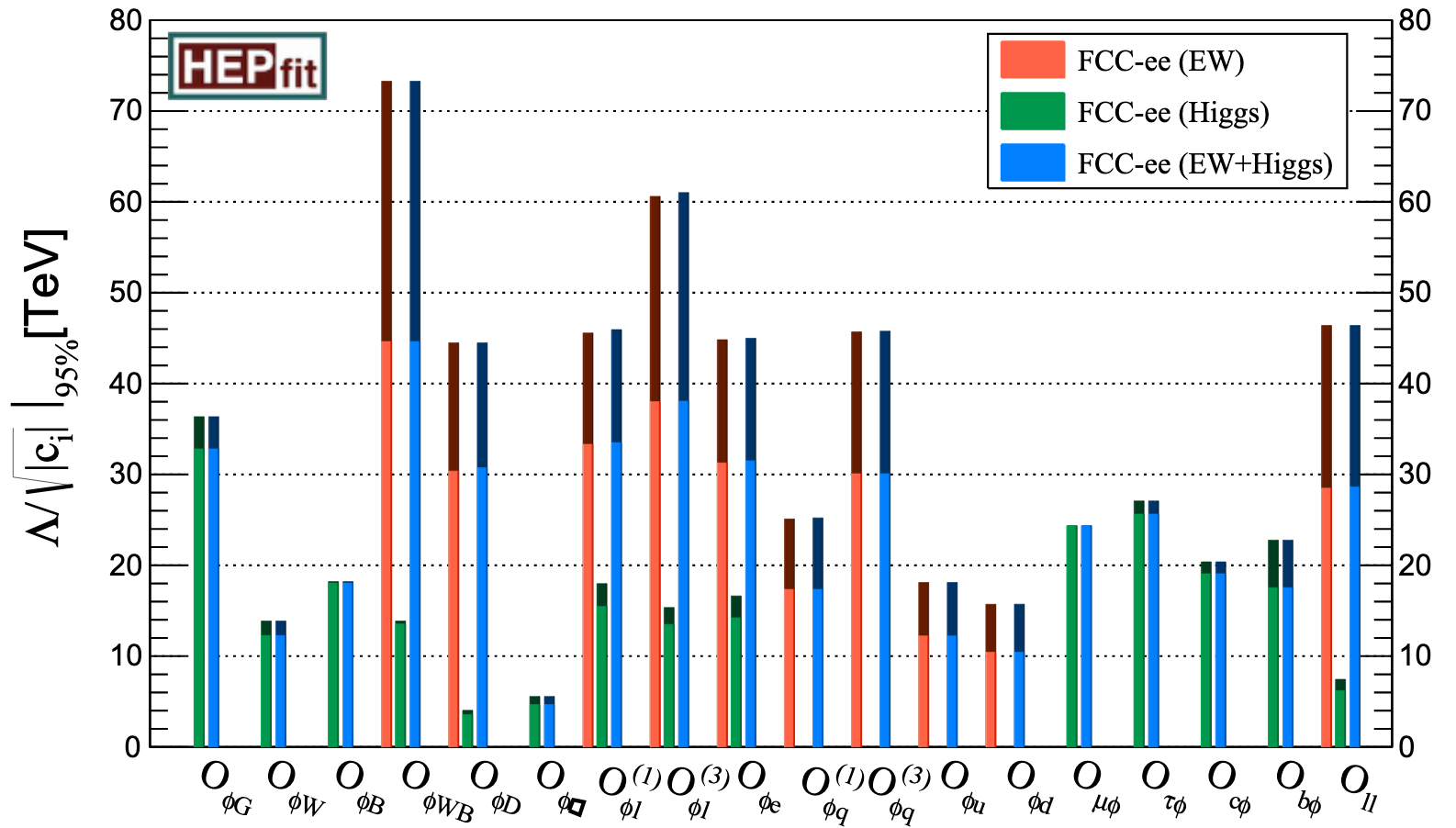}
\caption{\label{fig:Jorge} \small Electroweak (red) and Higgs (green) constraints from FCC-ee, and their combination (blue) in a global EFT fit. The constraints are presented as the 95\% probability bounds on the interaction scale, $\Lambda/\sqrt{c_i}$,  associated  to  each  EFT operator. Darker  shades  of  each  colour indicate the results when neglecting all SM theory uncertainties}
\end{figure}

The complementarity between the Higgs and Electroweak observables has been investigated in the context of specific heavy new physics, with a global EFT fit, as shown in Fig.~\ref{fig:Jorge}. 
This figure, made for FCC-ee only, highlights the complementarity of Higgs and electroweak measurements; the need for more statistics for the (statistics-limited) Higgs measurements; the interest of the Electroweak measurements and of the improvement of the associated systematic uncertainties; and the large number of observables available at FCC-ee. Not all observables of Table~\ref{tab:EWPO} have yet been used in this fit, and flavour observables have not been considered. 

Dedicated analysis of the pattern of deviations for specific models of new physics will be necessary to fully explore the ability of FCC-ee to identify or restrict the origin of one or several experimental deviation(s) from the SM predictions. The effects of a heavy ${\rm Z}^\prime$ gauge boson provide an illustrative example of complementarity, analysed in Ref.~\cite{Benedikt:2651299} for a specific Higgs composite model.
The precise measurements at and around the Z pole would be sensitive to such a new object by ${\rm Z/Z}^\prime$ mixing or interference, while measurements at  higher energies would display increasing deviation from the SM in the dilepton, diquark or diboson channels. The combination of these two effects would provide a tell-tale signature and allow constraints on mass and couplings of this possible new object to be determined.  

\section{Opportunities: Flavours}  
A total of $7 \times 10^{11}$ $\rm b \bar b$ pairs, available with a sample of $5 \times 10^{12}$ Z decays promised by FCC-ee, provides many opportunities in flavour physics. 
The precisions of CKM matrix element measurements expected from LHCb and Belle2 will be challenged, and the search for unobserved phenomena will be pushed forward, such as CP-symmetry breaking in the mixing of beautiful neutral mesons~\cite{Benedikt:2651299}. 

In parallel, searches for rare decays make FCC-ee a direct discovery machine. Lepton-flavour-violating (LFV) Z decays; rare and LFV $\tau$ decays; searches for heavy neutral leptons; and rare b-hadron decays; have been explored in Ref.~\cite{Benedikt:2651299} as benchmark or flagship searches. They are illustrative of the unique potential of a high-luminosity Z factory. The 
discovery potential relies on the vertexing capabilities of the experiments, to take benefit of the boosted topologies at the Z energy, but is mostly limited 
by the statistical size of the event sample.  

A minimum of $5 \times 10^{12}$  Z decays has been shown to be necessary to make, for example, a comprehensive study of the rare electroweak penguin transition $\rm b \to s \tau^+ \tau^-$~\cite{Kamenik:2017ghi}. For example, about 1000 events with a reconstructed $\rm \overline{B}^0 \to K^{\ast 0} \tau^+\tau^-$ are expected in such a sample. Should the current ``flavour anomalies"~\cite{Graverini:2018riw} persist, the study of b-hadron decays involving $\tau$'s in the final state is required to sort out possible BSM scenarios. If these flavour anomalies do not survive LHCb and Belle2 future scrutiny, the study of Z couplings to third-generation quarks and leptons still constitute an excellent opportunity to unravel BSM physics. 

Tau physics at the Z pole provides, still today, the most powerful tests of lepton universality and of the PMNS matrix unitarity. The increase of statistics by five orders of magnitude with respect to LEP should allow much improvement of these constraints, by combining measurements of the tau lifetime, the tau mass, and the leptonic branching ratios. 
These measurements provide important constraints on e.g. light-heavy neutrino mixing. Similar improvements are expected from the better precision of the  invisible Z decay width measurement~\cite{Benedikt:2651299,Abada:2019lih}. 

The exploration of the flavour physics program of FCC-ee is certainly in its infancy and can be expected to develop strongly in the future~\cite{Amhis:2021cfy}.

\section{Opportunities: QCD}

The $3.5 \times 10^{12}$ hadronic Z decay expected at FCC-ee also provide precious input for comprehensive QCD studies at the Z pole~\cite{Jorgen}. In particular, the determination of strong coupling constant $\alpha_{\rm S}(m_{\rm Z}^2)$, from the ratio $R_\ell$ of the Z hadronic width to the Z leptonic width, greatly benefits from the available statistics. With the present errors estimates, an experimental uncertainty on $\alpha_{\rm S}(m_{\rm Z}^2)$ of $0.00015$ can be contemplated, with further improvement being actively planned for the next round of detector studies. A similar figure can possibly be obtained from tau decays, or from the measurements of the hadronic and leptonic decay branching ratios of the W boson~\cite{Gomez-Ceballos:2013zzn,dEnterria:2016rbf,enterria-snow}, copiously produced with FCC-ee operating at larger centre-of-mass energies. 

The energy evolution of event shapes and fragmentation functions also provide powerful tests of QCD in $\rm e^+ e^-$ collisions. The wide range of energies covered by FCC-ee, which can start at centre-of-mass energies as low as 30\,GeV~\cite{Banfi-snow}, will provide yet another and completely independent determination of the strong coupling constant with a precision in the $10^{-4}$. 

Finally the Higgs run of FCC-ee contains an original gem: the possibility of clearly separating 
the $\rm H\rightarrow g g$ decay mode of the Higgs boson from the other hadronic decays expected to be completely dominated by $\rm b\bar{b}$  and $\rm c \bar{c}$ decays, could provide a unique laboratory to study the gluon fragmentation in a context unaffected by colour connection to quarks. 

\section{Opportunities: Direct observation of Feebly Coupled Particles}

\begin{figure}[!ht]
\centering
\includegraphics[width=0.70\textwidth]{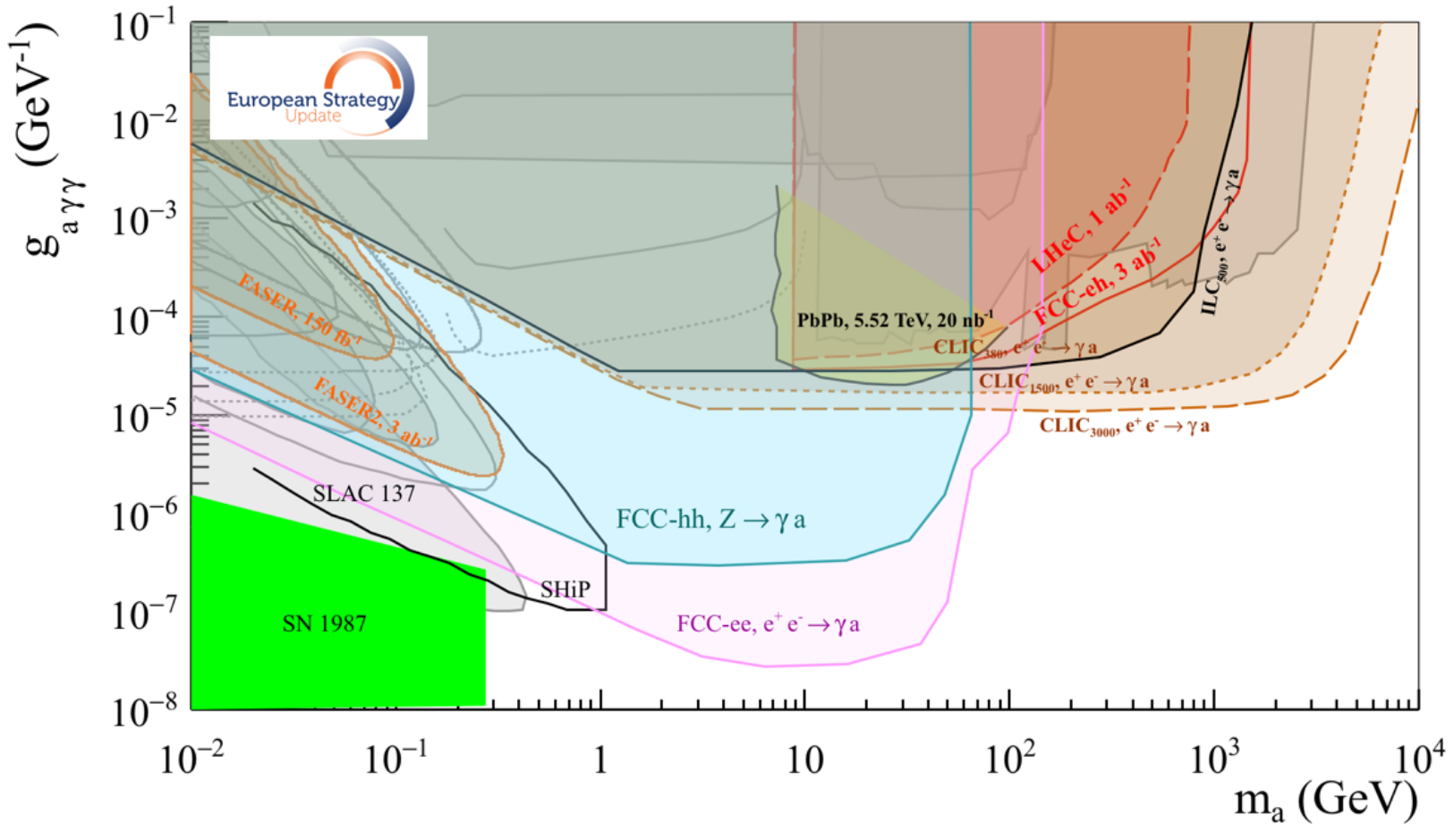}
\caption{\label{fig:ALPS} \small Expected sensitivity to axion-like particles in various future facilities. The reach of FCC-ee is down to very small couplings in Z decays, while the reach of linear colliders is at higher masses for somewhat larger couplings. From Ref.~\cite{Heinemann:2691414}  }
\end{figure}

Direct observation of new particles might require diverse searches and colliders to cover many orders of magnitude of coupling strength and mass scales. Feebly interacting particles such as  axion-like particles, or dark photons, with masses smaller  than the Z mass, may have the best odds of being found at the intensity frontier, i.e., with FCC-ee running at the Z pole, while high-energy machines might see them if their couplings are larger. The case of the axion-like particles is illustrated in Fig.\ref{fig:ALPS}. and demonstrates the high sensitivity of the  Z factory FCC-ee, as well as its complementarity with the FCC-eh and with a high-energy linear $\rm e^+ e^-$ collider. 

\begin{figure}[!ht]
\centering
\includegraphics[width=0.38\textwidth]{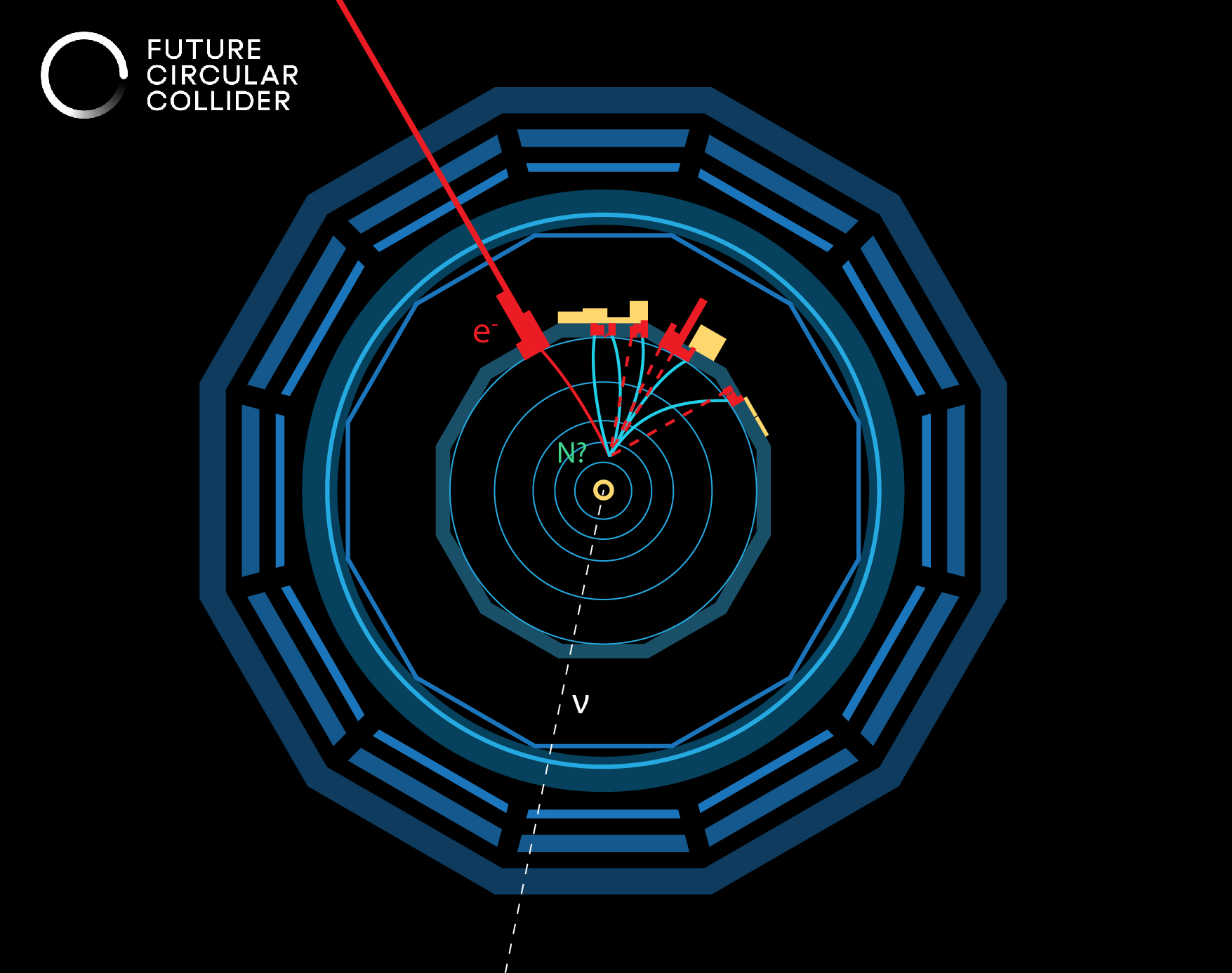}
\includegraphics[width=0.55\textwidth]{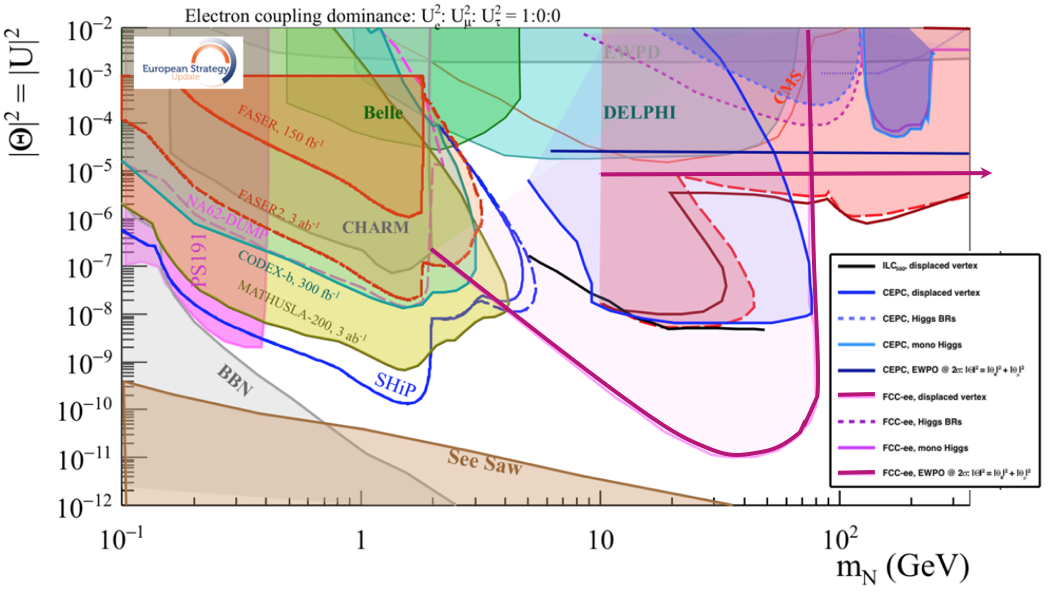}
\caption{\label{fig:RHnu} Left: sketch of the production of a Heavy Neutral Lepton at FCC-ee, $\rm e^+ e^- \rightarrow \nu N $ followed by the decay $\rm N \rightarrow e^- W^{*+}$ at about 1\,m from the interaction point. Right: Expected sensitivity to Heavy Neutral Leptons (a.k.a. Right Handed Neutrinos) in various future facilities. The reach of FCC-ee is for very small heavy-light mixing angle in Z decays, almost down to the see-saw limit; it is complemented up to very high masses (60\,TeV or more) for heavy-light neutrino mixing larger than $10^{-5}$ by constraints from Electroweak and tau decay precision measurements. From Ref.~\cite{Heinemann:2691414}.}   
\end{figure}

Another well-motivated example of new physics is provided by neutrinos. Many neutrino mass models naturally predict the existence of heavy neutrino states, called Heavy Neutral Leptons (HNL, mostly ``right-handed'' or ``sterile'') which mix with the known light, active neutrinos with a typical mixing angle  $|\theta_{\rm \nu N}|^2 \propto  m_{\nu}/m_{\rm N}  $. Since both light and heavy neutrino masses are unknown, a rather large range of mixing angles should be explored.   These scenarios  have several possible consequences: {\it (i)} the direct observation of a long-lived HNL in Z, W, and Higgs decays and in tau, b- or c-hadron semi-leptonic decays, both mass and mixing sensitive; {\it (ii)} the mixing of the light neutrinos with heavier states, which leads to a violation of the SM relations in EWPOs; the corresponding sensitivity only depends on the mixing angle, and extends to very high masses; {\it (iii)} the violation of lepton universality in $\tau$, b or c-hadron decays at the Z factory; {\it (iv)} a smaller-than-expected Z invisible decay width; and {\it (v)} a lepton-number violation can also result from Heavy-Neutral-Lepton production or exchange in high-energy processes at a hadron collider or a high-energy ${\rm e^- e^-}$ collider. The most sensitive tests {\it (i)} and {\it (ii)} for masses above $m_{\rm N} \geq 10$\,GeV are performed at FCC-ee, as shown on Fig.~\ref{fig:RHnu}, both for a possible direct observation, or for a well defined pattern of SM deviations in EW and HF observables.

\section{New Challenges}
Reaching experimental and theoretical systematic uncertainties commensurate to the statistical precision of the many measurements feasible at the FCC-ee requires careful preparation of the detector concepts, possibly of the mode of operation, and of theoretical developments. To this effect a certain number of benchmark measurements~\cite{Azzi:2021ylt} have been defined encompassing those listed in Table~\ref{tab:EWPO}. A repository of the Snowmass2021 documents describing them can be found in Ref.~\cite{LOIrep}. 

\subsection{Detector requirements}

The many opportunities of physics measurements and searches at FCC-ee will lead to many detector requirements, which may not be possible to assemble in a single or even two detector concepts. A complete set of such specifications and their impact on the physics measurements will be one of the main deliverables of the feasibility study. A first version is described in Refs.~\cite{PerezFCCPhys4,Mogens-ECFAinpt202102}. The following gives a brief idea of the type of requirements that are under study. 

\begin{itemize}
\item
{\bf Higgs and top physics}
A set of requirements for Higgs and top physics at 240 and 340--365\,GeV, arising in particular from the desired performance of particle-flow reconstruction, flavour tagging, or lepton momentum, have been established by past linear collider studies. They need to be  adapted to take into account the more gentle FCC-ee operating conditions, the smaller beam-pipe radius, and the smaller maximum operating energy~\cite{Azzi:2021gwg}. Additional requirements arise, related to the need for a more accurate Higgs boson mass determination~\cite{Azzurri:2021nmy} prior to the $s$-channel Higgs production run, and for a more accurate ZH cross-section measurement, to enable a precise determination of the  Higgs self-coupling. 
The $s$-channel Higgs production also leads to demanding requirements on the centre-of-mass energy monochromatisation, while keeping a high luminosity and requesting the most sensitive analysis to separate Higgs decay from the huge background of $\epem$ annihilation events~\cite{dEnterria:2021xij}. 

\item
{\bf TeraZ challenges}
The high luminosity and large event rate at the Z pole (over 100~kHz), to which simulated data need to be added, turn into to considerable challenges for data taking, storage and processing. The Z lineshape determination, which is based on cross-section measurements as a function of the centre-of-mass energy for hadronic and leptonic Z decays of the Z, requires extremely accurate mechanical construction of the luminometer, as well as precise knowledge of the central detector (tracker and/or calorimeter) acceptance, for the di-lepton and di-photon events (and to a lesser extent, for hadronic events). The point-to-point centre-of-mass-energy uncertainties of the resonance scan, which can be verified in particular by means of the muon pair invariant mass reconstruction, are most relevant for the Z width and the forward-backward asymmetries. This sets stringent constraints on the stability of the momentum reconstruction over time and scan points~\cite{Maestre:2021zvq}. 

\item 
{\bf W mass determination} The W mass can be obtained from a scan of the WW pair threshold. A precision of 0.5\,MeV seems a priori more demanding on the centre-of-mass energy calibration, and on the theoretical understanding of the cross section, than on the detector itself. The requirements on jet angular calibrations and the possible dependence of the precise knowledge of the composition of hadrons in these jets, however, should probably be revisited. Similarly, requirements on lepton angle, energy reconstruction, and energy scale, should be established~\cite{Azzurri:2021yvl}.

\item
{\bf Flavour challenges}
The availability of $7 \times 10^{11}$\,${\rm \ b\bar{b}}$ events, paired with a formidable vertexing ability, leads to unprecedented b- and c-tagging performance. The resulting expected statistical uncertainties on the flavour EWPOs, such as $R_{\rm b,c}$ or $A_{\rm FB}^{\rm b,c}$, allow for much larger improvements with respect to the LEP measurements than other EWPOs. In addition, the rich FCC-ee flavour program can be fully exploited only if the detector is equipped with hadron identification covering the effective momentum range at the Z resonance~\cite{WilkinsonFCCPhys3}, and electromagnetic calorimetry with an energy resolution of $3-5\%/\sqrt{E}$~\cite{AleksanFCCPhys4}.

\item
{\bf Tau physics}
The $\tau$ measurements listed in Table~\ref{tab:EWPO} (lifetime, mass, and branching fractions) have the potential to achieve a determination of the Fermi constant at the level of a few $10^{-5}$. These measurements provide some of the most demanding detector requirements, on momentum resolution (for the mass measurement and the LFV search), on the knowledge of the vertex detector dimensions (for the tau lifetime), and $\rm e/\mu/\pi$ separation over the whole momentum range (for the leptonic branching ratios). In addition, the $\tau$-based EWPOs, $ R_{\tau}^{\rm Z}$, $A_{\rm FB}^{\rm pol,\tau}$ and $P_\tau$, as well as the detailed study of the hadronic spectral functions, require fine granularity and high efficiency in the tracker and electromagnetic calorimeter~\cite{Dam:2021pnx}.  

\item
{\bf Rare Processes}~\cite{Chrzaszcz:2021nuk} 
The current HNL search strategy is based on a rather conservative signal selection, requiring in particular a HNL decay less than 1\,m away from the interaction point. The discovery region is just shy of the see-saw limit (Fig.~\ref{fig:RHnu}). It would be possible to extend it by detecting HNL decays further away from the IP, e.g., by making use of the large amount of cavern space surrounding the detector~\cite{Chrzaszcz:2020emg}. Other feebly interacting particles, decaying into non-pointing or delayed photons, pose a different set of challenges, which could possibly benefit from high precision timing.  
\end{itemize}

Integrating all these (initial) detector requirements in one or even several detectors will be a considerable challenge, commensurate with the unique set of measurement opportunities and with the discovery potential offered by FCC-ee.  A refined list of detector requirements is being derived from a number of case studies inspired from a set of physics benchmark measurements at FCC-ee, as described in Ref.~\cite{Azzi:2021ylt}. The performance of a number of options for the various detector components of possible future detectors:  calorimeters~\cite{R1calo}; tracking and vertex detectors~\cite{R1tracking}; muon detectors~\cite{R1muondet}; luminometers~\cite{Dam:2021sdj}; and particle identification devices~\cite{Wilkinson:2021ehf}; are being studied to understand how they could meet the requirements. 

\subsection{The case for four interaction points}
\label{section:4IP}

One of the great advantages of circular $\epem$ colliders is the possibility of serving several interaction points with a net overall gain both in integrated luminosity and luminosity per MW. It is expected that a layout with four interaction regions could be designed with a total facility luminosity larger by a factor 1.7 than that of the scenario studied so far, with only two interaction points. 

The rich menu of the FCC-ee physics possibilities combines a set of precision measurements, feebly-coupled particle sensitivity, and rare process studies, which will challenge and shape particle physics for many decades. There is a genuine chance for new physics discovery, and such a result could have a significant impact on FCC-hh detector design and physics outcome.  The key of success is a blend of high luminosity, redundancy, and careful preparation of detector setups. Many measurements are statistics limited and immediately benefit from four interaction points. Several key physics targets are tantalisingly close with the present two-IP setup, as indicated with the Higgs self-coupling and the search for heavy neutral leptons. Having four IPs allows for a range of detector solutions to cover all FCC-ee physics potential opportunities. 

The studies already at hand indicate that the variety of detector requirements (placed, e.g., on the electromagnetic calorimeter) may not be satisfied by one or even two detectors: the constraints on high precision, high granularity,  high stability, geometric accuracy. particle identification, cost, etc. may be too many to solve at once. Finally, experience from LEP taught us that different detector solutions are invaluable in uncovering hidden systematic biases and avoid conspiracy of errors, while providing an attractive challenge for all skills in the field of particle physics: detector technology, design and  R\&D, computing, software, analysis, and theory. 

\subsection{Theoretical Challenges}

The FCC-ee physics program presents a number of key theoretical challenges, discussed in Chapter III of this EPJ+ special issue~\cite{Heinemeyer:2021rgq}. Generally speaking,  
the aim is either to provide the tools to compare experimental observations to theoretical predictions at a level of precision similar or better than the (statistical) experimental uncertainties, some of which are listed in Table~\ref{tab:EWPO}; or to identify the additional calculations, tools, observables, or experimental inputs that are required to achieve this level of precision. Another precious line of research to be followed jointly by theorists and experimenters is to identify observables, or ratios of observables, for which experimental and/or theoretical uncertainties can be reduced. Finally, both for motivational purposes and for prioritzation, the relative impact of the various measurements on the search for new physics should be evaluated. The theoretical work motivated by the FCC program can be organised as follows.
\begin{itemize}
    \item  Calculation of QED (mostly), EW, and QCD corrections to (differential) cross sections, needed to convert experimental measurements back to ``pseudo-observables'': couplings, masses, partial widths,  asymmetries,  etc.  that are close to the experimental measurement (i.e.,  the relation between measurements and these quantities does not alter too significantly the possible ‘new physics’ content).
    Appropriately accurate event generators are essential for the implementation of these effects in the experimental procedures.
\item  Calculation  of  the  pseudo-observables  with  the  precision  required  in  the  framework  of  the  Standard Model so as to take full advantage of the experimental precision.
\item 
Identification of the limiting issues and, in particular, the questions related to the definition of parameters, in particular, the treatment of quark masses and, more generally, QCD objects.
\item 
An investigation of the sensitivity of the proposed (or new) experimental observables to the effect of new physics in a number of important specific scenarios. This essential work must be done at an early stage, before the project is fully designed, since it potentially affects the priorities for detector concepts and for the running plan.
\end{itemize}
An important community of theorists has already risen to the precision challenges, especially at the Z peak~\cite{Blondel:2018mad}. An evaluation of the options for the path  ahead can be found in Refs.~\cite{Blondel:2019qlh,Freitas:2019bre,Heinemeyer:2021rgq}. 

\section{Conclusions}
\label{section:conclusion}

The alignment of stars that led, in 2011/2012, to the concept of a 100\,km-class electron-positron collider and, in 2020, to the visionary update of the European Strategy for Particle Physics endorsing the FCC feasibility study as a top priority for CERN and its international partners~\cite{CERN-ESU-015}, provides the HEP community with a powerful tool of investigations. 
Such a machine offers ideal conditions (luminosity, centre-of-mass energy calibration, possibly monochromatisation) for the study of the four heavy particles of the Standard Model with a flurry of opportunities for precision measurements, searches for rare or forbidden processes, and the possible discovery of feebly coupled particles. The FCC-ee is a also perfect springboard for a 100\,TeV hadron collider, for which it provides a great part of the infrastructure. The complementary and synergistic physics programs of these two machines offer a uniquely powerful long-term vision. 

The work for the particle physicists is thus clearly cut out: design the experimental setup and prepare the theoretical tools that can, demonstrably, fully exploit the FCC-ee capabilities. Experimentation at FCC-ee is both relatively easy and extremely demanding. The experimental conditions are clean, with essentially no pile-up, well-defined and controllable centre-of-mass energy, benign beamstrahlung and synchrotron radiation effects. The challenges arise from the very richness of the program, with $5.10^{12}$ Z (including $1.5\, 10^{11}$ muon pairs and tau pairs and $7(6).10^{11}$ of b (c) quark pairs),  $3.10^8$ W pairs, and $10^6$ ZH and ${\rm t\bar t}$ events. Matching the experimental and theoretical accuracy to the statistical accuracy, and the detector configuration with the variety of channels and discovery cases, might appear to be mission impossible. Of course, nothing will be impossible for the group of creative enthusiasts, who are poised to tackle this exciting set of challenges.

%
%
%

\bibliographystyle{myutphys}
\bibliography{references}

\section*{\small Data availability}

{\small \it Raw data were generated at CERN. Derived data supporting the findings of this study are available from the corresponding author upon request.}
\end{document}